\documentclass[aps,prl,twocolumn,showpacs,preprintnumbers,amsmath,amssymb]{revtex4-1}
\usepackage{graphicx}
\usepackage[usenames,dvipsnames]{xcolor}
\usepackage{lineno}
\usepackage[caption=false]{subfig}
\DeclareMathOperator{\Tr}{Tr}

\begin{document}
%\preprint{DRAFT}
\title{Probing the conformal Calabrese-Cardy scaling with cold atoms}
\author{J. Unmuth-Yockey$^1$}
\author{Jin Zhang$^{2}$}
\author{P. M. Preiss$^{3}$}
\author{Li-Ping Yang$^4$}
%\author{Authors}
\author{S.-W. Tsai$^2$}
\author{Y. Meurice$^1$}
%\email[]{yannick-meurice@uiowa.edu}
\affiliation{$^1$ Department of Physics and Astronomy, The University of Iowa, Iowa City, IA 52242, USA }
\affiliation{$^2$ Department of Physics and Astronomy, University of California, Riverside, CA 92521, USA}
\affiliation{$^3$ Physikalisches Institut, Heidelberg University, 
69120 Heidelberg, Germany}
%\affiliation{Other Adresses}
\affiliation{$^4$ Department of Physics,Chongqing University, Chongqing 401331, China}
\definecolor{burnt}{cmyk}{0.2,0.8,1,0}
\def\lt{\lambda ^t}
\def\note{note}
\def\beq{\begin{equation}}
\def\enq{\end{equation}}

\date{\today}
\begin{abstract}

We demonstrate that current experiments using cold  bosonic atoms trapped in one-dimensional optical lattices  and designed to measure the second-order R\'{e}nyi entanglement entropy  $S_2$,  can be used to verify detailed predictions of 
conformal field theory (CFT) and estimate the central charge $c$. We discuss the adiabatic preparation of the ground state at half-filling where we expect a CFT with $c=1$. This can be accomplished with a very small hopping parameter $J$, in contrast to existing studies with density one where a much  larger $J$ is needed. We provide two complementary methods to estimate and subtract the classical entropy generated by the experimental preparation and imaging processes. We compare numerical  calculations for the classical $O(2)$ model with a chemical potential on a  1+1 dimensional lattice, and the quantum Bose-Hubbard Hamiltonian implemented in the experiments. $S_2$ is very similar for the two models and follows closely the 
Calabrese-Cardy scaling, $(c/8)\ln(N_s)$, for $N_s$ sites with open boundary conditions, provided that the large subleading corrections are taken into account.

\end{abstract}


\maketitle

The concept of universality provides a unified approach to the critical behavior of 
lattice models studied in condensed matter, lattice gauge theory (LGT) and 
experimentally accessible systems of cold atoms trapped in optical lattices. Conformal field theory (CFT)
\cite{1987gauge,difrancesco} offers many interesting examples of universal behavior that can be observed for lattice models in two \cite{Belavin:1984vu,Friedan:1983xq,Dotsenko:1984if}, three \cite{PhysRevD.86.025022}, and four \cite{DeGrand:2010ba,Kuti:2015awa} dimensions. 
Practical simulations for these models unavoidably involve a finite volume that breaks explicitly 
the conformal invariance. However, this symmetry breaking follows definite patterns dictated by 
the restoration of the symmetry at infinite volume and allows us to identify the universality class. 
In view of the rich collection of interesting CFTs, it would be highly desirable to study their universality classes using quantum simulations.
In order to start this ambitious program, one needs a simple concrete example to demonstrate the feasibility of the idea.

In this Letter, we propose to use the setup of ongoing cold atom experiments to quantum simulate the $O(2)$ model with a chemical potential and 
check the predictions of CFT for the growth of the entanglement entropy with the size of the system corresponding to the universality class of the superfluid (SF) phase. 
The $O(2)$  model is an extension of the Ising model where the spin is allowed to move on a circle,  
making an angle $\theta$ with respect to a direction of reference.  This model can be used to describe easy plane ferromagnetism and the compactness of $\theta$ leads to topological configurations called vortices. Their unbinding 
provides a prime example of a Berezinski-Kosterlitz-Thouless transition \cite{b,kt} in a way that has also been advocated to apply for gauge theories near the boundary of the conformal window \cite{Kaplan:2009kr}.
When space and Euclidean time are treated isotropically, 
this model has important common features with models studied numerically in LGT to describe relativistic systems in the continuum limit. 
Quantum simulating this model and studying experimentally the CFT predictions  would be a crucial first step towards applying similar methods for LGT models. 

%In these examples, the conformal symmetry is explicitly broken by the lattice regularization and only %emerges in the continuum and infinite volume limits. Identifying the underlying conformal symmetry %through well-understood 
%symmetry breaking patterns in numerical or experimental  simulations involving finite lattices is an %important tool to 
%explore new universality classes. 

In the following, we show that these goals can be achieved by measuring the entanglement entropy of a simple Bose-Hubbard (BH) model in a very  specific region for the adjustable couplings.
The entanglement entropy measures the correlations between degrees of freedom 
in different regions of a system, is an important tool \cite{vidal03} in assessing the phase structure and the finite-size scaling. 
For a CFT in one space and one time (1+1) dimension, 
the ground state entanglement entropy increases logarithmically with the spatial volume of the system and its subsystems  
\cite{vidal03,korepin04,Calabrese:2004eu,Calabrese:2005zw,PhysRevLett.104.095701,unusualcardy}. Using  basic CFT results, Calabrese and Cardy 
\cite{Calabrese:2004eu} established  that the coefficient of proportionality is in general the  central 
 charge multiplied by a rational number depending on the type of entropy and the boundary conditions (CC scaling). 
The central charge, denoted $c$, is of primordial importance in CFT.
%plays a crucial role in the construction of the unitary representations of the conformal algebra, 
It characterizes the universality class and  is present in a variety of physical observables \cite{difrancesco,Calabrese:2004eu}. 

%In view of the rich collection of interesting CFTs in 1+1 dimensions, it would be highly desirable to study their universality classes using quantum simulations. 

It has been proposed to use a quantum gas microscope to   study the second-order R\'enyi entropy $S_2$ of one-dimensional  fermionic Hubbard chains  \cite{twinfermions,PhysRevB.88.155103} at half(quarter)-filling which seem consistent with $c=1(2)$. 
Recently, manipulations of small one-dimensional chains of cold bosonic $^{87}$Rb atoms trapped in optical lattices have allowed experimental measurements of $S_2$ \cite{Islam2015,Kaufman2016} using a beam splitter method  \cite{PhysRevLett.93.110501, PhysRevLett.109.020505}. 
In these experiments, the  SF phase is  
reached by  increasing  the hopping parameter  $J$ to values having the same order of magnitude as the onsite energy $U$.  It is important to realize that  in  order to see  a clear correspondence  between  the BH  and $O(2)$ model,  
one needs $J/U\ll 1$. Some examples are given below. However, the feasibility of the measurements is restricted by the fact that very small values of $J$ can be problematic because of disorder or finite-temperature effects. We argue that a reasonable compromise is to take $J/U\simeq 0.1$. In this regime, a detailed study  \cite{Jdraft,SM} shows that the finite size scaling is easier to resolve near half-filling  \footnote{For bosons, {\it half-filling} means twice more sites than particles, while for spin-1/2 fermions it means one particle per site.}.
% and where a  SF  phase 
%with  significant values for $S_2$ can be  reached at small $J$. 

Experimental measurements have been performed for small chains of four \cite{Islam2015} and six \cite{Kaufman2016} atoms and only slightly larger sizes are expected to be within experimental reach in the near future \footnote{In the discussion we focus on  experiments with 16 sites or less.}. 
An important feature of $S_2$ with open boundary conditions \cite{Jdraft,SM} is that the subleading corrections are large and decay slowly with $N_s$ (see Eq. (\ref{eq:fit}) below).  Knowing these corrections is essential to extract the leading CC scaling using $N_s$ accessible in experiments.

The details of the connection between the $O(2)$ and BH models are discussed in \cite{PhysRevA.90.063603}. 
The von Neumann entanglement for the isotropic model has been calculated numerically \cite{PhysRevE.93.012138} using tensor Renormalization Group (TRG) methods \cite{prd88,PhysRevA.90.063603}.  
In order to connect the $O(2)$ model with quantum simulators, it is possible to take a highly anisotropic limit of the transfer matrix where the time becomes continuous and we can identify 
a quantum Hamiltonian \cite{PhysRevD.17.2637,RevModPhys.51.659,M.Fisher.2004,sachdev2011,PhysRevA.90.063603} :
\begin{equation}
	\hat{H}=\frac{U}{2}\sum_x \hat{L}_x^2-\tilde{\mu}\sum_x 		\hat{L}_x-2J\sum_{\left<xy\right>}\cos(\hat{\theta}_x-			\hat{\theta}_y) \ ,
\label{eq:rotor}
\end{equation}
with $[\hat{L}_x,{\rm e}^{i\hat{\theta} _y}]=\delta_{xy}{\rm e}^{i\hat{\theta} _y}$. These commutation relations can be approximated 
with finite integer spin  \cite{PhysRevA.90.063603}. 
In the following, we use  
the  spin-1 and spin-2 approximations for numerical calculations. % when $\lambda<1$. 
Quantum simulators involving two species of bosonic atoms have been 
proposed for the spin-1 approximation \cite{PhysRevA.90.063603,PhysRevD.92.076003}. 
This effort is directly related to recent attempts (for recent reviews see Refs. \cite{Tagliacozzo:2012vg,wiesereview,Zohar:2015hwa}) to develop quantum simulators for models studied in LGT. 
Note that the non-zero eigenvalues of $\hat{L}$ come in pairs with opposite signs. When $\tilde{\mu}=0$, the sign of these eigenvalues plays no role and there is an exact invariance under the charge conjugation which implies the existence of anti-particles. 

\begin{figure}[h]
\includegraphics[width=8.6cm]{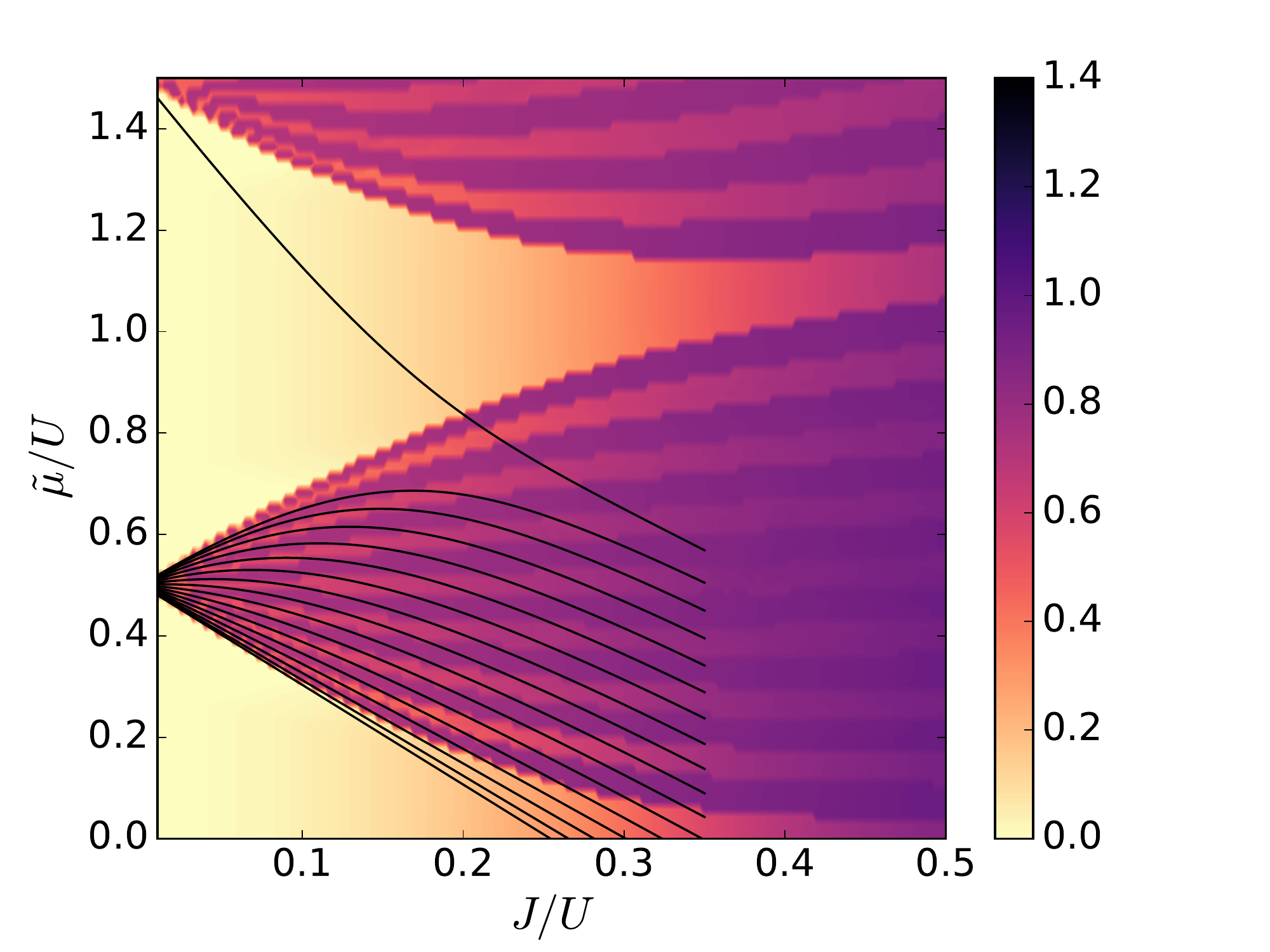}
\caption{$S_2$ for $O(2)$ with $N_s=16$ and OBC. Laid over top are the BH boundaries between particle number sectors.}
\label{fig:motttosf}
\end{figure}

On the other hand, when $\tilde{\mu}$ is large and positive, the states with negative eigenvalues play a minor role in numerical calculations. If we omit these states, we can replace $\hat{L}_x$ by the occupation number $n_x$ and ${\rm e}^{i\hat{\theta} _x}$ by the creation operator $a^\dagger_x$ in Eq. (\ref{eq:rotor}). We then obtain the 
simple BH Hamiltonian: 
\begin{equation}
\hat{H}=\frac{U}{2}\sum_x n_x(n_x-1)-J\sum_{x}(a^{\dagger}_xa_{x+1}+h.c.) .
\end{equation}
This approximate correspondence already discussed in the literature \cite{M.Fisher.2004,sachdev2011} is supported by  results presented below. 
In the following, we focus on the region of the phase diagram where 
$\tilde{\mu}\simeq U/2\gg J$ illustrated in Fig. \ref{fig:motttosf}. In this regime, 
the particle occupancies 0 and 1 dominate for BH (hard core limit) and there is an {\it approximate} correspondence  with the spin-1/2 XX model 
which is integrable and has a central charge $c=1$ \cite{vidal03,korepin04}. 

Recent cold atom experiments \cite{Islam2015,Kaufman2016} have  measured the second-order R\'enyi entropy 
\begin{equation}
S_2(\mathcal{A})\equiv -\ln(\Tr(\rho_\mathcal{A}^2)) \ ,
\end{equation}
for a variety of subsystems $\mathcal{A}$ and open boundary conditions (OBC). The reduced density matrix $\hat{\rho}_\mathcal{A}$ is obtained by tracing over the complement of $\mathcal{A}$. 
CFT provides severe restrictions on the dependence of $S_2$ on the size of the system and the subsystem
\cite{Calabrese:2004eu,Calabrese:2005zw,unusualcardy,PhysRevLett.104.095701,1742-5468-2011-01-P01017}. 
In the following, we restrict ourselves to systems with an even number of sites and a subsystem $\mathcal{A}$ of size $N_s/2$. 
Fits of other subsystems will be discussed in  \cite{Jdraft,SM}.
Fig. \ref{fig:motttosf} displays $S_2$ for $N_s=16$ as a function of $J/U$ and the chemical potential. The lower (upper) light part is the Mott phase with particle density $\lambda=$ 0 (1), and the 15 plateaus corresponding to the particle number  sectors 1, 2, \dots, $N_s-1$ in the SF phase in between are 
visible. Its boundaries $\tilde{\mu}/U=1/2\pm2J/U$ at small $J$ follow from a perturbative calculation and are consistent with Refs. \cite{PhysRevB.46.9051,monien98} for BH at larger $J$ .
In the following, we focus on the half-filling region which is more or less horizontal 
in the SF region and can be reached numerically at arbitrarily small $J/U$. 

Since existing experiments only allow a very limited number of sites, 
it is crucial to take into account subleading corrections. Using existing results
\cite{Calabrese:2004eu,Calabrese:2005zw,unusualcardy,PhysRevLett.104.095701,1742-5468-2011-01-P01017}
for subsystems of size $N_s/2$, we consider the form:
\begin{equation}
	S_{2}(N_{s}) = K + A \ln(N_{s}) +
    \frac{B \cos\left(\frac{\pi N_{s}}{2} \right)}{(N_{s})^{p}} +
    \frac{D}{\ln^{2}(N_{s})}, 
    \label{eq:fit}
\end{equation}
where  $K, \  A,\  p, \  B,\ $ and $D$ are fitting  parameters. For OBC, the CC scaling predicts $A=c/8$. 
In order to verify this  prediction, we have calculated $S_2$ at half-filling for $J/U=0.1$ 
for the two models considered with the Density Matrix Renormalization Group (DMRG) method  \cite{PhysRevLett.69.2863,PhysRevLett.75.3537} using the ITensor C++ library \footnote{version 2.7.10, http://itensor.org/}. For the $O(2)$ model, the results were cross-checked \cite{Jdraft,SM} with TRG methods \cite{prd88,PhysRevA.90.063603,PhysRevE.93.012138}.

\begin{figure}[ht]
\includegraphics[width=8.6cm]{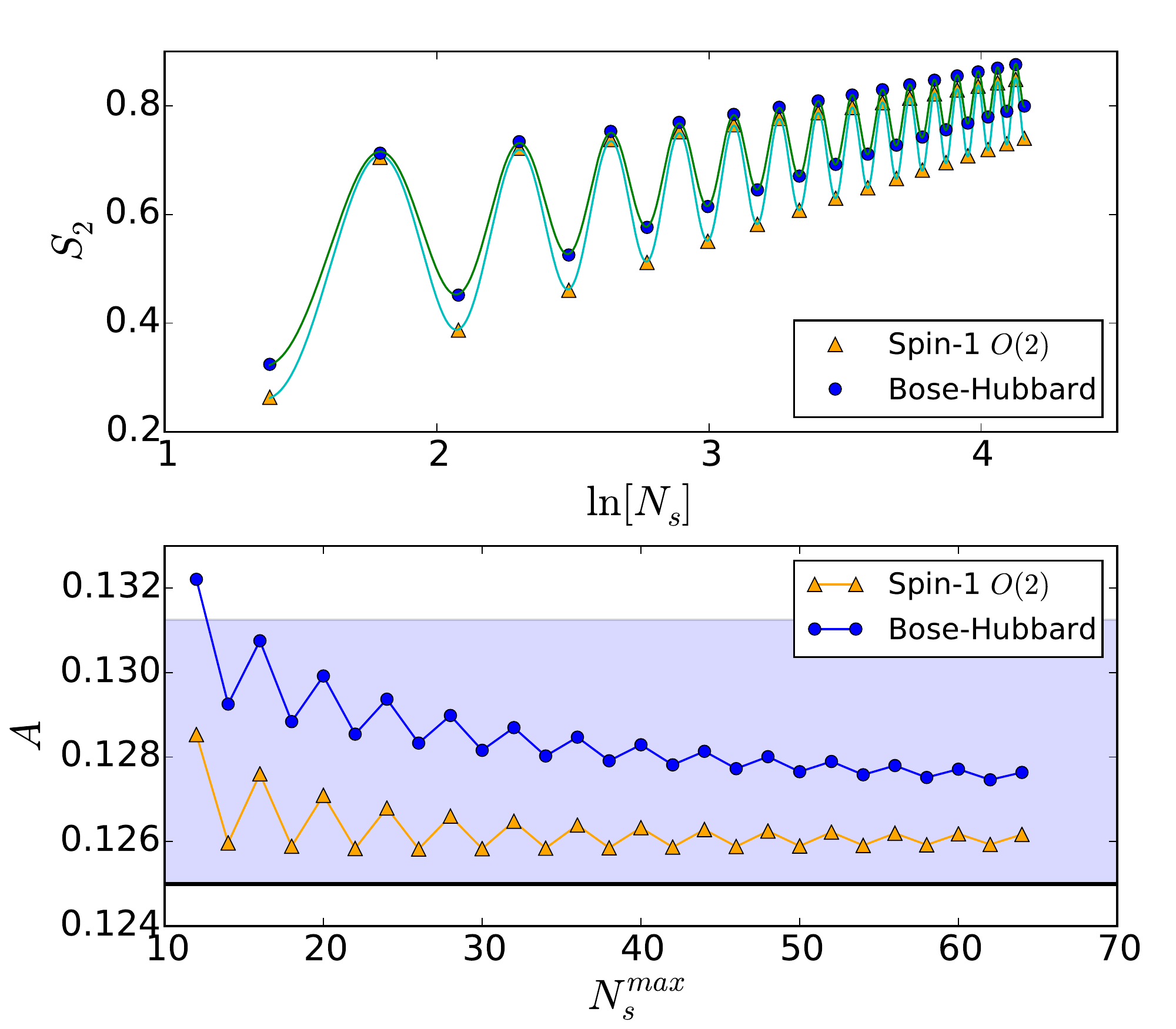}
\caption{(top) $S_2$ at half-filling with OBC for  $O(2)$ and BH with $J/U=0.1$.  The solid lines are the fits for BH and $O(2)$. (bottom) Values of $A$ as a function of the maximal value of $N_s$ used in the fit, the band represents a positive departure of 5 percent from the expected value 0.125. }
\label{fig:RE}
\end{figure}

If we use the numerical data for $N_s$ up to 64, we obtain $A=$ 0.1263 for $O(2)$ and 0.1278 for BH which is close to the CC prediction 0.125 for $c=1$. 
The difference between the two models can be reduced significantly by decreasing $J/U$, which also brings $A$ closer to 0.125 \cite{Jdraft,SM}. In order to test the predictive ability of the fit for smaller spatial sizes we have reduced 
the maximal value $N_s^{max}$ of $N_s$ from 64 to smaller values, down to 12. The results for $S_2$ and $A$ are 
shown in Fig. \ref{fig:RE} which suggests that the estimates converge slowly to the CFT value as $N_s^{max}$ increases. It has also been noticed that if  $J/U$ is increased to $J/U\simeq 0.3$ for BH, the sign of $D$ changes in a way that seems almost independent of the other subleading corrections used. In this region of parameters, the periodic corrections are smaller which may facilitate the estimate of $c$, however the close connection with the $O(2)$ model is lost.

\begin{figure}[ht]
\includegraphics[width=8.2cm]{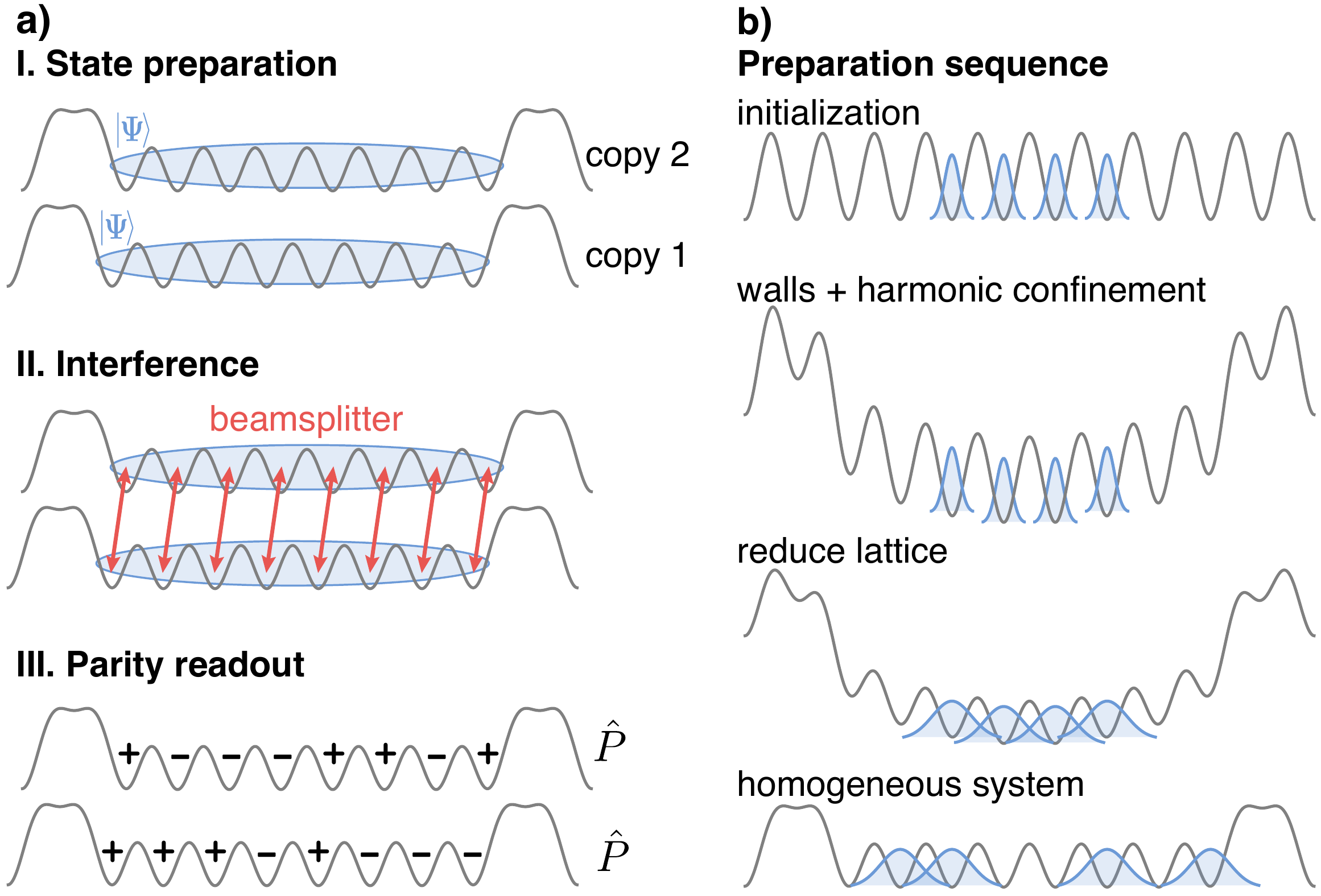} 
\caption{Measuring entanglement entropy in optical lattices. a) Two copies of a quantum state $| \Psi \rangle $ interfere under a beamsplitter operation, and site-resolved number measurements reveal the local parity $\hat{P}$ and entanglement entropy. b) Proposed state preparation for BH systems at 
half-filling, here for 4 atoms on 8 sites. Particles indicated by wavefunctions (blue online) are initialized in a deep optical lattice, where the local environment can be shaped via harmonic confinement and sharp features projected with a spatial light modulator. As the lattice depth is reduced, the particles delocalize but are confined by repulsive walls.}
\label{fig:experiment}
\end{figure}

We now proceed to explain the proposed experimental setup. 
We consider an optical lattice experiment with single-particle resolved readout and local manipulation of the optical potential, similar to Ref. \cite{Islam2015}.
In the experiment, two copies of the one-dimensional many-body state of interest are prepared in adjacent rows of an optical lattice,  and global and local R\'{e}nyi entropies can be measured by a beamsplitter operation implemented via a controlled tunneling operation between the two copies (Fig. \ref{fig:experiment}a). The parity of the atom number in one copy after the beamsplitter operation gives access to the quantum mechanical purity \cite{PhysRevLett.109.020505}.

BH systems with tunable parameters $U$ and $J$ and well-defined particle number are realized in current experiments with one particle per site. Fig. \ref{fig:experiment}b shows a proposed scheme to achieve half-filling at $J/U\approx 0.1$: $N_p$ bosons are initialized in the Mott regime $J \ll U$, as in current experiments. A superimposed harmonic confinement as well as two sharp, confining walls separated by $N_s$ sites ensure that the system remains in its ground state as the optical lattice depth is adiabatically reduced to achieve the desired $J/U$. The harmonic confinement is then removed to realize a homogeneous system with hard wall boundary conditions at half-filling. For system sizes considered here, this scheme should allow adiabatic preparation of the ground state with available experimental tools. Alternatively, techniques based on optical superlattices may be able to prepare lattice ground states at half-filling \cite{Trotzky2012}.

After preparing twin tubes with half-filling in their ground state and applying the beamsplitter operation, one can measure 
the number of particles modulo 2 at each site $x$ of a given copy ( $n_x^{copy}$) \cite{Islam2015}, and use the result
\cite{PhysRevLett.109.020505}:
\begin{equation}
\exp (-S_2) = \Tr(\rho _\mathcal{A}^2)= \langle(-1)^{\sum_{x\in \mathcal{A}} n^{copy}_x}\rangle \ ,
\label{eq:parities}
\end{equation}
to calculate $S_2$. 

The probability for parity $(-1)^{n_x}=\pm 1$ is $(1\pm \exp (-S_2))/2$. As 
$S_2$ increases, more cancellations occur and one needs on the order of $\exp(2S_2)$ measurements to overcome the fluctuations. From Fig. \ref{fig:RE}, and assuming $N_s$ to be less than 16 (i. e., less than 8 particles at half-filling with an entropy per particle of order 0.05), the maximal measured $S_2$ is less than 1.1. For  $\mathcal{N}$ independent measurements, we find that the  statistical error is 
\begin{equation}
\sigma_{S_2} = \sqrt{(e^{2S_2}-1)/\mathcal{N}} \ .
\end{equation} For the maximal value $S_2=1.1$, it takes about 800 measurements to reach $ \sigma_{S_2} \simeq 0.1$. Due to the logarithmic growth of $S_2$, the number of measurements only needs to increase like 
$N_s^{1/4}$ to maintain a desired accuracy, which is not a prohibitive growth.

In addition to the statistical errors, one needs to take into account that finite temperature as well as preparation and 
manipulation errors contribute a classical entropy $S^{class.}$. Assuming that this classical entropy is linear in the number of particles in the system, it can be estimated by making use of an approximate particle-hole symmetry: near half-filling, $S_2(N_s)$ of the ground state is in good approximation symmetric in the particle number about $N_p=N_s/2$. By measuring $S_2^{exp.}(N_s)$ for a range of particle numbers in the vicinity of $N_s/2$, the excess classical entropy per particle in the experiment can be determined. Subtracting this estimate of the classical entropy from the experimentally measured $S_2^{exp.}$ gives a corrected estimate of the ground state entanglement entropy $S_2^{corr.}$, which we compare to CFT via Eq. (\ref{eq:fit}). For the system sizes considered here, deviations from an exact particle-hole symmetry are small  and exhibit a regular behavior  at  zero and  finite temperature  \cite{Cardy:2014jwa}. Understanding and fitting these effects  is important  to get  estimates  of  $S_2^{corr.}$ with  errors  less  than 0.02 \footnote{Jin Zhang et al. , work in progress.}.

\begin{figure}[h]
\includegraphics[width=8.6cm]{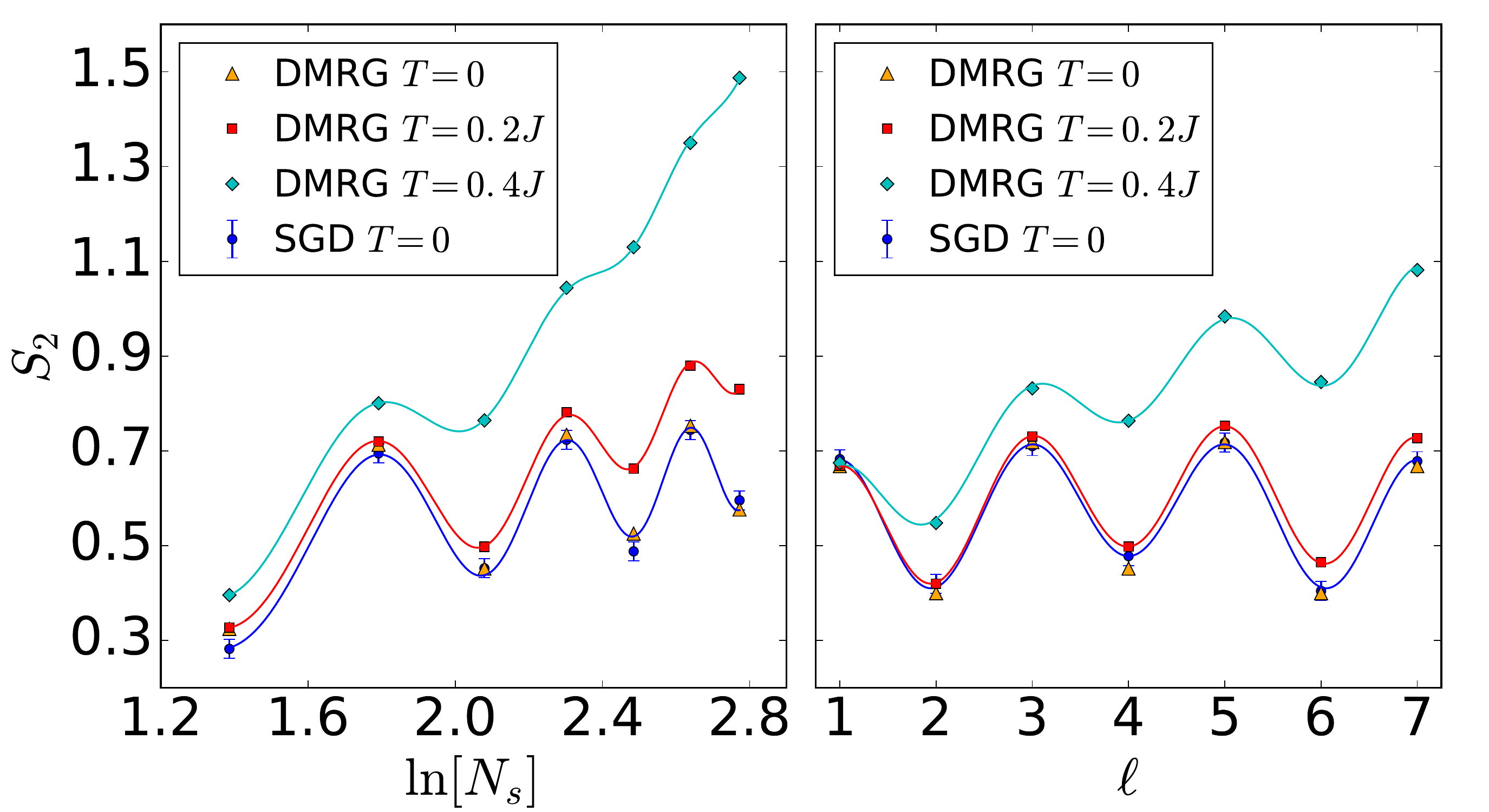}
\caption{\label{fig:bh4exp}$S_2$ at half-filling for BH with $J/U = 0.1$ (triangles, orange online) and SGD with random Gaussian fluctuations with $\sigma_{S_2} = 0.02$ (circles, blue online).  (left): vs. $\ln(N_s)$ for a  subsystem of size $N_s/2$ with the solid line corresponding to a fit of the SGD from Eq. (\ref{eq:fit}).
(right): vs. the subsystem size $\ell$ for $N_s=8$; the solid line corresponds to a  fit of the SGD using the formulas of Ref. \cite{PhysRevB.83.214425}. Same  quantities  for  $T=0.2J$ (squares, red online) and $T=0.4J$  (diamonds, cyan online).}
%\label{fig:bh4exp}
\end{figure}

In order to give an idea of possible experimental outcomes, we have numerically studied the sensitivity of the fit results of 
Eq. (\ref{eq:fit}) to statistical errors in the measured values of $S_2$. By repeatedly fitting synthetically generated data (SGD) with Gaussian noise on $S_2$ of magnitude $\sigma_{S_2}$ as illustrated in Fig. \ref{fig:bh4exp} (left), we find that it translates into errors of the fit approximately as $\sigma_A=3.1\sigma_{S_2}$ for a global fit of the central charge involving data up to $N_s=16$. To reach a statistical uncertainty in $A$ comparable to systematic errors of the order 0.02, the statistical error on $\sigma_{S_2}$  has to be on the order of 0.005.  

Alternatively, we can try to fit $S^{class.}$. For this purpose, we have considered the  
finite temperature ($T$) effects for $T=0.2J$ and $0.4J$ in Fig. \ref{fig:bh4exp} (left).  Remarkably, these  effects 
can be  fitted by adding only one term linear  in $N_s$.  If  $S^{class.}$ generated  during the experiment  follows  this linear behavior,  it  may be used to  determine  some effective temperature.  Note that the finite temperature  effects become  more  important  as we decrease $J$ \cite{SM}.

So far we have only used the values of $S_2$ corresponding to a subsystem of size $N_s/2$. CFT also provides prediction for arbitrary subsystem sizes $\ell$ with $1\leq \ell \leq N_s-1 $ which are described in the SM \cite{SM}.
The large oscillations when $\ell$ is changed for $N_s=8$ are shown in Fig. \ref{fig:bh4exp} (right). Finite-$T$ effects can be fitted with a single 
additional term linear in $\ell$ \cite{SM}. 
Importantly, the experimental measurements 
of the parities at each site shown in Eq. (\ref{eq:parities}) allow us to calculate $S_2$ for {\it all} possible subsystems without extra measurements. Estimates of $c$ from numerical calculations at fixed  $N_s$ fits in other models  
have up to 20 percent errors \cite{mazza15,Jdraft,SM}.
Knowing $S_2$ for all the subsystems also allows us to 
calculate the mutual information \cite{twinfermions,  Islam2015}, where the $S^{class.}$ contributions cancel.

New directions should be pursued. Half-filling initial states can also be obtained by a sudden expansion.  The presence of additional approximate conserved charges makes the thermalization non-trivial and interesting \cite{PhysRevA.82.063605,daley14,2014arXiv1408.1041S,PhysRevB.88.235117}. The possibility of revivals in the time-dependent $S_2(t)$ for time scales of the order of 200 ms for $J/U=0.1$, a duration about 10 times longer than current experiments \cite{Kaufman2016}, is under study.
The techniques discussed here for the bosonic case can also be applied to Fermi-Hubbard systems \cite{twinfermions}, for which optical lattice experiments with single-site resolution are rapidly becoming available \cite{Haller2015,PhysRevLett.115.263001,PhysRevLett.114.193001,PhysRevLett.114.213002}.
It would be desirable to develop specific procedures to study models with other values of $c$ (Ising, $Z_N$ clock, Potts) or with $O(3)$ symmetry with a chemical potential, which have a similar phase diagram \cite{falko3}, and could be quantum simulated \cite{Laflamme:2015wma}. More insight on conformal symmetry could be gained by studying particle number fluctuations 
\cite{chinscale,PhysRevLett.108.116401,PhysRevLett.115.035302}. 
The entanglement entropy can also be calculated in 
pure gauge theories using standard Monte Carlo methods \cite{buisu2}. 
Methods for calculating the entanglement entropy in the presence of fermion determinants  have been designed on the lattice \cite{drut15} and in the continuum \cite{kl11}. 

In conclusion, we have shown that the simple BH model which is implemented in current experimental measurements of $S_2$  can be used as a quantum simulator for the classical $O(2)$ 
model with a chemical potential. We showed that the region of the phase diagram near half-filling and small $J/U$ offers rich possibilities that complement the existing experiments at unity-filling and larger $J/U$ \cite{Islam2015,Kaufman2016}.
The changes in $S_2$ due to the size of the system or the subsystem show strong periodic oscillations which are of the same order of magnitude as the average $S_2$ for $N_s\leq 16$.  
We  provided complementary methods  to  estimate and subtract  $S^{class.}$ from $S^{exp.}_{2}$.
Existing experiments could immediately confirm the periodic patterns found in the numerical calculations and fits. Accurate determination of $c$ would require larger statistics or a 
suitable use of the complete information about the subsystems. 
Conformal symmetry connects disparate physical systems from condensed matter and LGT. While this equivalence is usually apparent in theory in the thermodynamic limit, we have shown that the basic equivalence 
between the BH model and the classical $O(2)$ model can already be identified in present cold-atom experiments. Our proposed method could enable the first direct verification of conformal scaling in an experimentally accessible system.

\vskip20pt

\begin{acknowledgments}
Acknowledgments. We thank M. C. Banuls, I. Bloch, I. Cirac, M. Greiner, A. Kaufman, G.  Ortiz, J. Osborn, H. Pichler and 
P. Zoller for  useful suggestions or comments. 
This research was supported in part  by the Department of Energy
under Award Numbers DOE grant DE-FG02-05ER41368, DE-SC0010114 and DE-FG02-91ER40664, the NSF under grant DMR-1411345 and by the Army Research Office of the Department of Defense under Award Number W911NF-13-1-0119.  
L.-P. Yang was supported by Natural Science Foundation for young scientists of China (Grants No.11304404) and Research Fund for the Central Universities(No. CQDXWL-2012-Z005). P.M.P.  has received funding from the European Union's Horizon 2020 research and innovation programme under the Marie Sklodowska-Curie grant agreement No 706487.  Parts of the  numerical calculations were done  at the Argonne  Leadership  Computational Facilities. Y.M. thanks the Focus Group Physics with Effective Field Theories of the Institute for Advanced Study, Technische Universit{\"a}t M{\"u}nchen, and the workshop on ``Emergent properties of space-time" at CERN for hospitality while part of this work was carried out and the Institute for Nuclear Theory for motivating this work during the workshop ``Frontiers in Quantum Simulation with Cold Atoms".
\end{acknowledgments}
\input{bib.bbl}
%\bibliography{centralmacbib2.bib}
%\end{document}
\newpage

\begin{figure}
  \includegraphics[width=8.5cm]{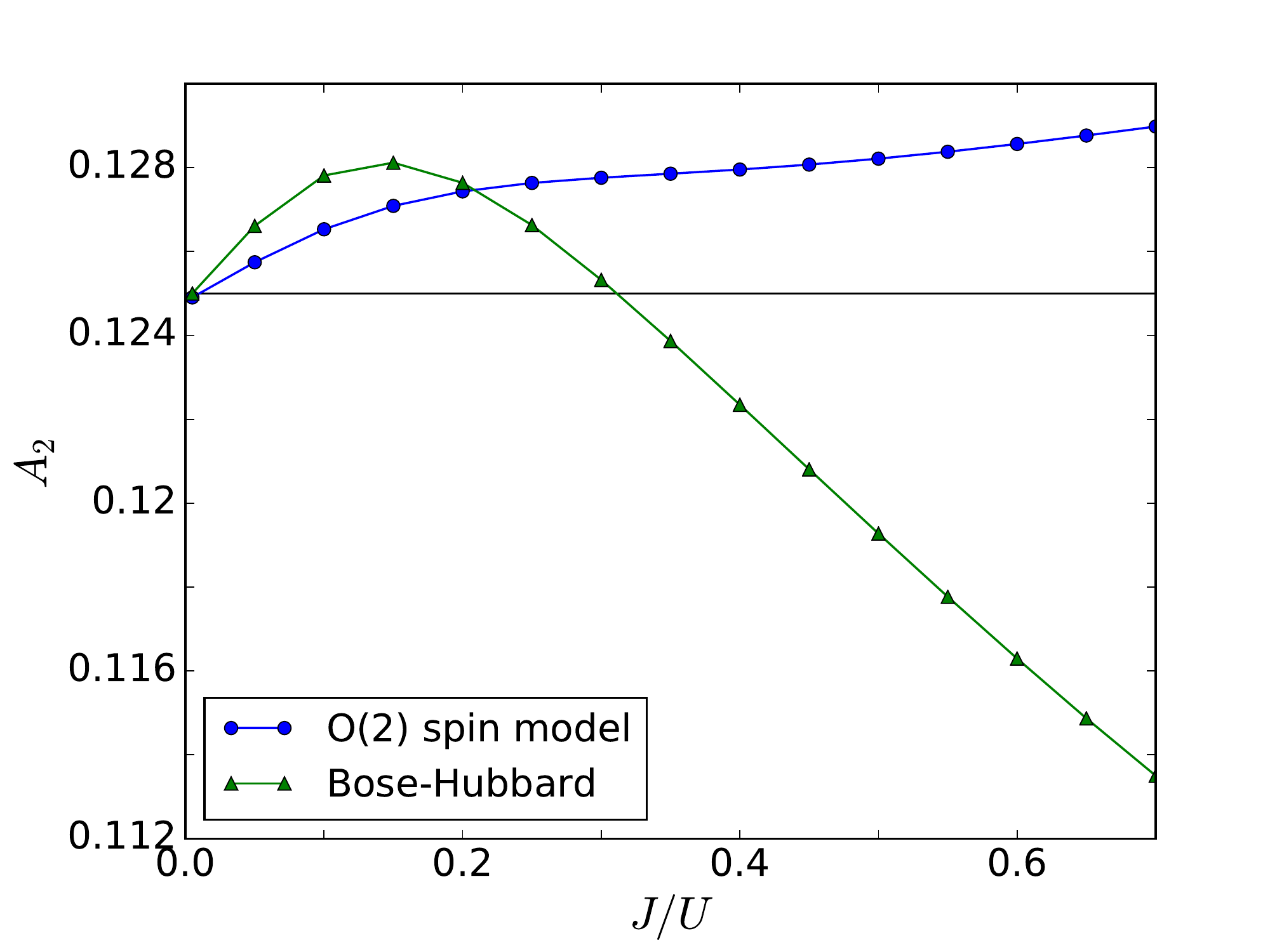}
  \caption{The $A$ values from fits to $S_{2}$ with open boundary conditions for Bose-Hubbard data (green triangles online) and $O(2)$ spin model data with a spin-1 truncation (blue circles online) at half-filling.  The horizontal line is the conformal field theory prediction.  The $A$ values were extracted from a fit to Eq. \ref{eq:cft-fit} with a correction term proportional to $1/\ln^{2}(N_{s})$.  The farthest point to the left is at $J = 0.005$, and the lines through the data are meant to guide the eye.}
      \label{fig:AvsJ}
	\end{figure}

\begin{center}
{\bf Supplemental Material}
\end{center}
This \emph{Supplemental material} aims to provide background calculations done for the Bose-Hubbard (BH) and $O(2)$  model that give empirical evidence to the claims made in the main text.  The main text put forward that in the right region of coupling space (i.e. small enough hopping $J$ relative to the on-site repulsion, $U$) at half-filling, it is possible to quantum simulate the $O(2)$ model using a single species Bose-Hubbard model, and in said simulation, measure the R\'{e}nyi entanglement entropy and subsequently the conformal central charge, $c$.
We considered the BH and $O(2)$ models at half-filling, however in this text the $O(2)$ model only makes an appearance in Fig \ref{fig:AvsJ}.

The choice for half-filling was made because of the following: Near the boundaries of the Mott insulator lobes with the superfluid phase the R\'{e}nyi entropy takes on a constant value of $\ln(2)$.  Between each lobe there are $N_{s}-2$ boundaries denoting the particle number in the chain, with an approximate symmetry around $N_{s}/2$ particles (i.e. half-filling for bosons).  This number has the clearest signal for the entanglement entropy (see Ref. \cite{Jdraft2}) and therefore all work here, and in the main text, was done at half-filling. 
In addition, all work here and in the main text was done with open boundary conditions.  However, while this is more easily realized in experiment, the second order R\'{e}nyi entropy with open boundary conditions has much larger fluctuations than the von Neumann entropy (See Ref. \cite{Jdraft2}).  

Here we first consider fits to the second order R\'{e}nyi entanglement entropy, $S_{2}$, across a range of $J/U$ values to better understand the dependence on $J/U$.  Next we consider fits made to data across the entire range of system sizes, $N_{s}$, and across all subsystem sizes, $l$.  These fits span the entire $N_{s}$-$l$ plane and aim to explain how we decided to specifically focus on the case $l = N_{s}/2$.  We report the average relative error for fits to large and small systems.  We mention some investigation into taking finite temperature effects into account in experiments.  Finally we discuss how we used synthetic Gaussian fluctuations to simulate errors on experimental data and estimate the error that would be incurred on measuring the central charge.

In order to understand how the fits were influenced by $J/U$, we considered fits of $S_{2}$ across a range of $J/U$ values.  These fits were done with DMRG data for $N_{s}=4$ up to $N_{s}=64$.  We fit the second order R\'{e}nyi entanglement entropy with a subsystem size of $l=N_{s}/2$  and compared the coefficient of logarithmic scaling, $A$, between the BH and $O(2)$ models for various finite-size correction terms.  The $A$ values as a function of $J/U$ can be found in Fig. \ref{fig:AvsJ} with a correction term proportional to $1/\ln^{2}(N_{s})$ which is predicted by conformal field theory (CFT) \cite{unusualcardy2}.  We considered additional corrections like $1/N_{s}$, $1/\ln(N_{s})$, etc \ldots  
We found some features were robust across different functional forms for the corrections.  For larger $J/U$ the BH $A$ values tended to decrease monotonically, while the $O(2)$ $A$ values seem to increase.  There appears to be a crossing between the BH data and the CFT prediction around $J/U \approx 0.35$.  However at small $J/U$, only the correction $\propto 1/\ln^{2}(N_{s})$ showed a tendency towards $A_{2} \approx 1/8$ as $J/U \rightarrow 0$, a soluble limit where $c=1$.  The choice for $J/U=0.1$ for quantum simulation is a nice compromise since ideally the smaller the $J/U$ the better for the mapping between BH and $O(2)$, however, for experimental purposes too small a value of J is inconvenient because of the associated long time scales and sensitivity to uncontrolled disorder, and finite temperature effects are larger at smaller $J/U$.

At this point, it's important to reiterate that calculations done in the rest of this text specifically refer to the BH model.  A deeper investigation into the $O(2)$ spin model can be found in Ref. \cite{Jdraft2}.

CFT gives predictions for the scaling of the R\'{e}nyi entropy as a function of subsystem size. Eq. \ref{eq:cft-fit} gives the leading order prediction along with a term believed to account for finite-size effects and parity oscillations \cite{PhysRevB}.  Importantly, the experimental measurements of the parities at each site shown in Eq. (5) in the main text %this is in the main text!!!!!
allow us to calculate $S_{2}$ for all possible subsystems without extra measurements.  The possibility of using these additional (but statistically correlated) results to reduce the overall statistical error on the estimate is under study.

\begin{figure}
    \includegraphics[width=8cm]{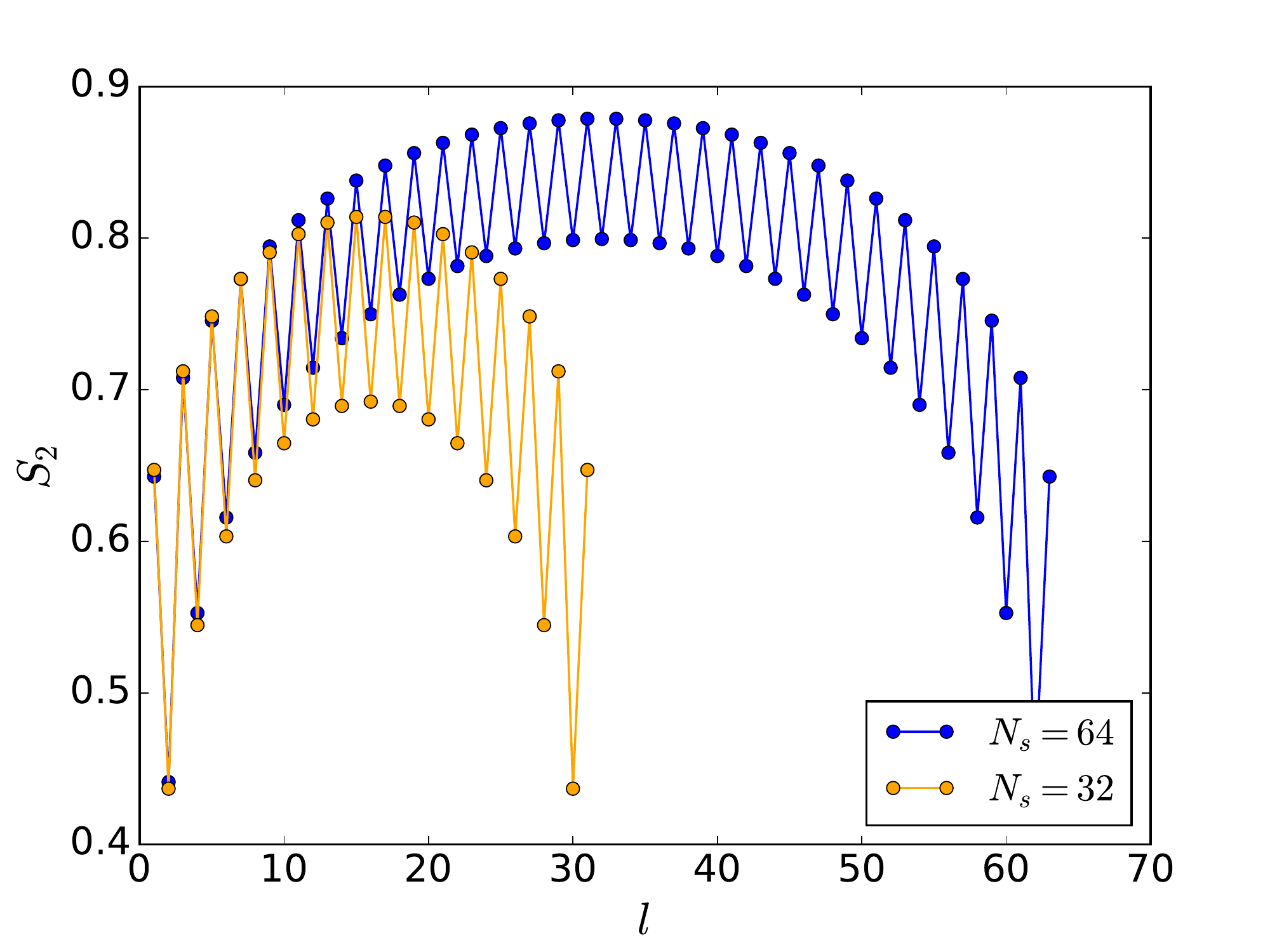}
    \caption{$S_{2}$ for the Bose-Hubbard model at $J/U=0.1$, with open boundary conditions. Here the data from $N_{s} = 32$ and $64$ is plotted together as an example of the oscillatory behavior of the data as a function of the subsystem size, $l$.}
    \label{fig:bh-32-64-data}
\end{figure}

We explored $S_{2}$ as a function of system size, $N_{s}$, and subsystem size, $l$, in the BH model. As an example we have the subsystem data plotted for two different system sizes, $N_{s} = 32$ and 64, in Fig. \ref{fig:bh-32-64-data}.
From the plot one can see that for small $l$ and $l \approx N_{s}$, the amplitudes are almost independent of $N_{s}$, however near $l \approx 12$ the data sets depart.  We attempted fits using
\begin{align}
\label{eq:cft-fit}
	S_{n}(N_{s}, l) &= A_{n} \ln \left\{ N_{s} \sin \left[  \frac{\pi l}{N_{s}} \right] \right\} + B \\
    &+ \frac{C}{N_{s}^{p_n}} \cos ( \pi l )
    \left| \sin \left[ \frac{\pi l}{N_{s}} \right] \right|^{-p_n} \nonumber \\
    &+ f_n(N_{s}, l) \nonumber
\end{align}
as well as including the order $1/l$ corrections \cite{PhysRevB.83.214425},
\begin{align}
\label{eq:xavier-fit}
    S_{n}(N_{s}, l) &= A_{n} \ln \left\{ \frac{4(N_{s}+1)}{\pi} \sin \left[  \frac{\pi (2l+1)}{2(N_{s}+1)} \right] \right\} + B \\
    &+ \frac{C}{N_{s}^{p_{n}}} \cos ( \pi l )
    \left| \sin \left[ \frac{\pi (2l+1)}{2(N_{s}+1)} \right] \right|^{-p_{n}} \nonumber \\
    &+ f'_{n}(N_{s}, l) \nonumber
\end{align}
where $f'_{n}$ and $f_{n}$ are correction terms.  
We performed global fits across the entire $N_{s}$-$l$ plane in order to take full advantage of the data.
% The absolute value of the difference between the fits and the data is plotted in Fig. \ref{fig:allpointsfit} with the correction $\propto 1/\ln^{2}(N_{s})$, although other corrections were tried.
% \begin{figure}
%     \includegraphics[width=0.5\textwidth]{BH-2fits-l-Ns.pdf}
%     \caption{The absolute value of the difference between BH data and fits using Eq. \ref{eq:xavier-fit} (left) and \ref{eq:cft-fit} (right) in the $N_{s}$-$l$ plane with $f$, $f' \propto 1/\ln^{2}(N_{s})$.}
%     \label{fig:allpointsfit}
% \end{figure}

\begin{figure}
\begin{tabular}{cc}
	\subfloat{\includegraphics[width=0.2\textwidth]{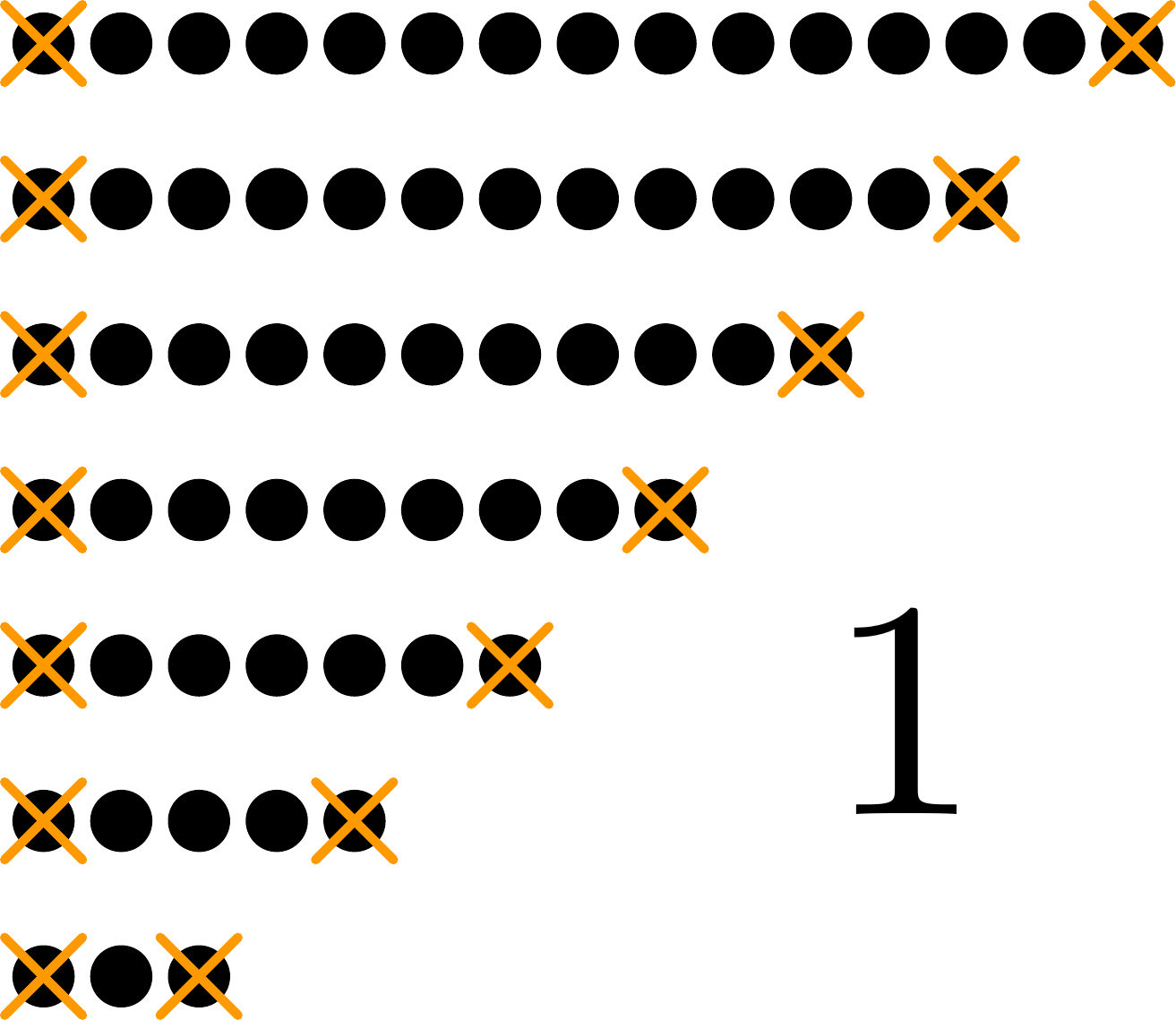}} &
	\subfloat{\includegraphics[width=0.2\textwidth]{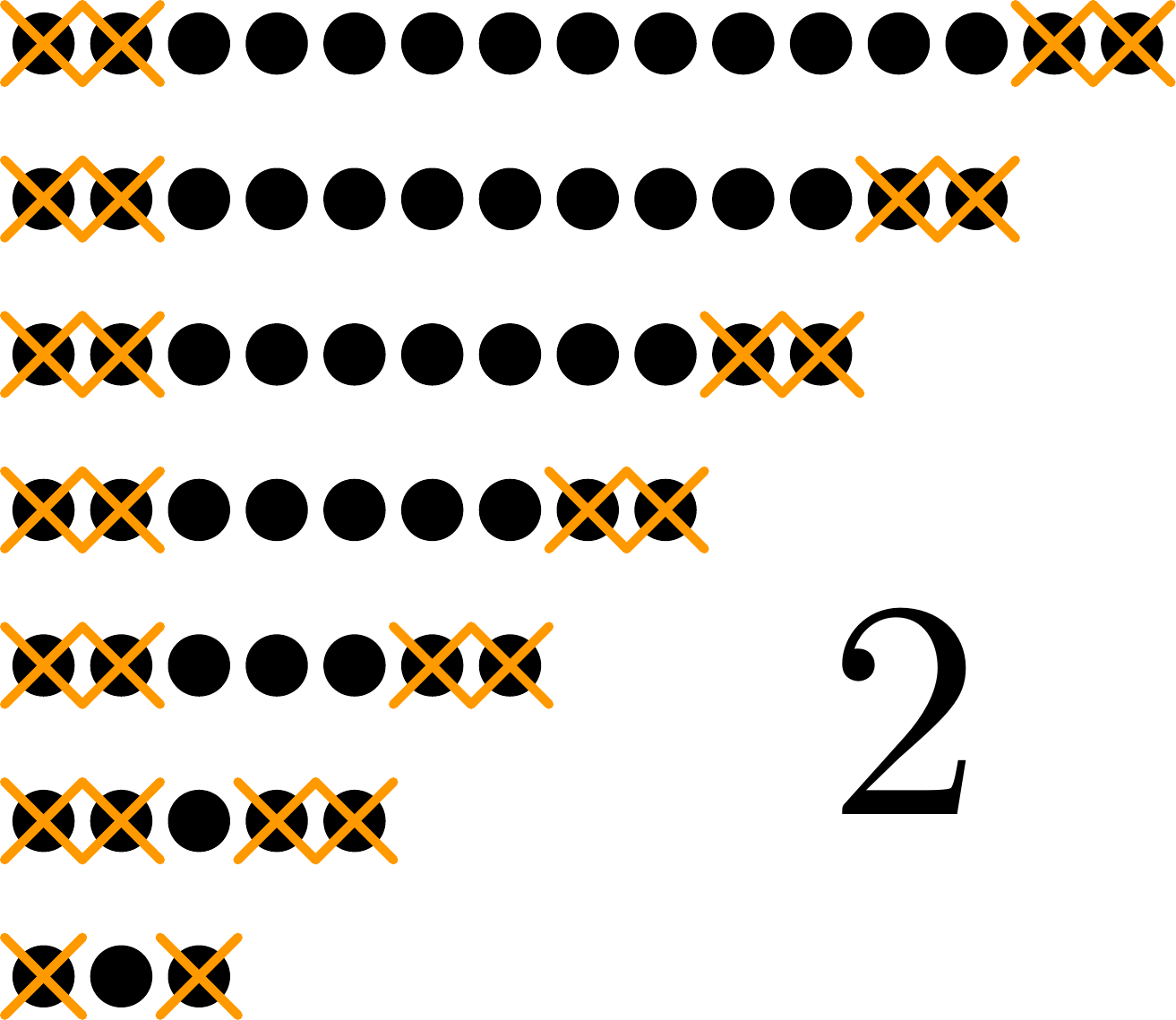}} \\
    \subfloat{\includegraphics[width=0.2\textwidth]{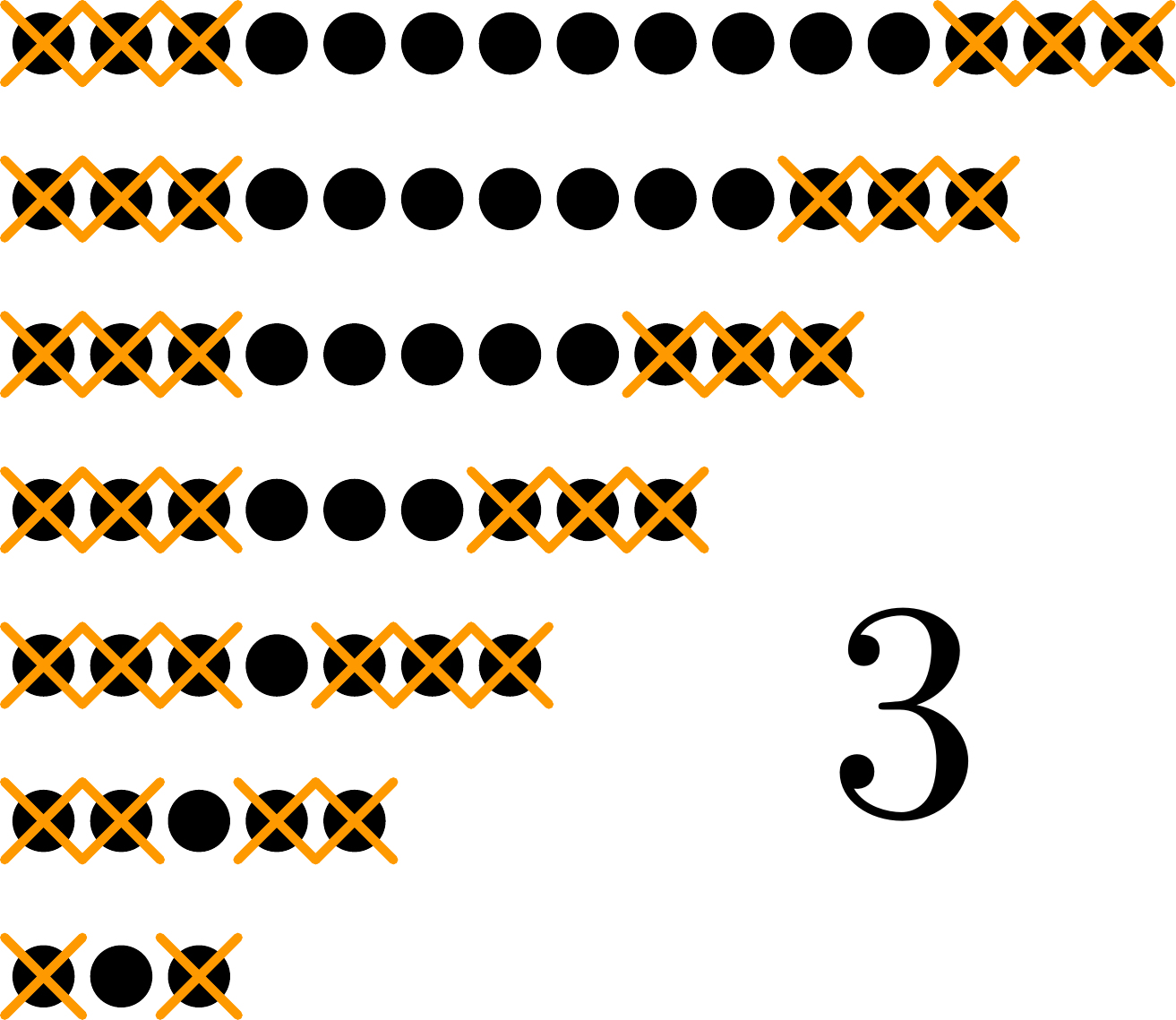}} &
    \subfloat{\includegraphics[width=0.2\textwidth]{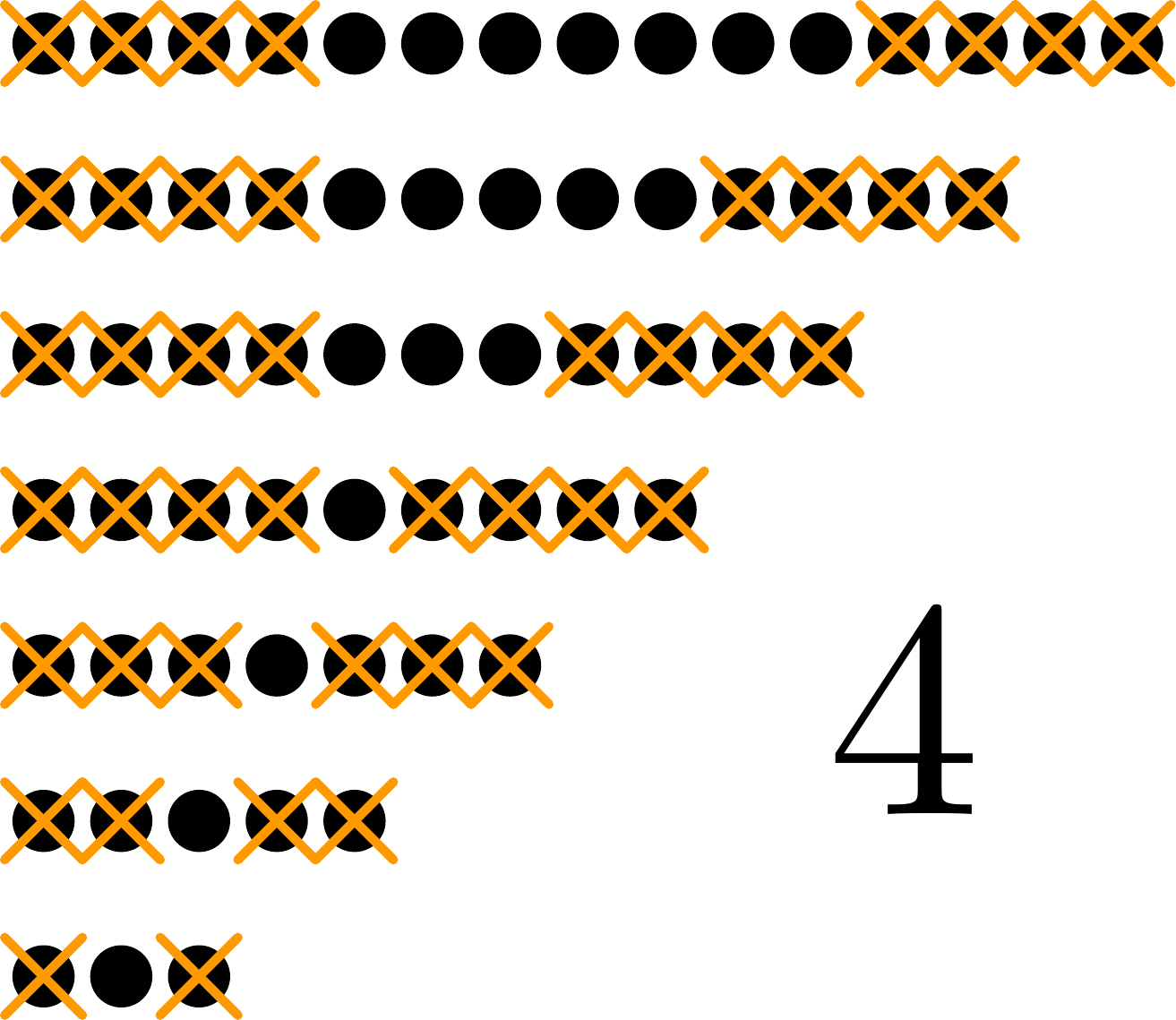}}
\end{tabular}
\caption{How points are removed when preforming a global fit to the entire data set to determine the influence of the boundary points on the average relative error.  Here the system size increases in the vertical direction ($N_{s}$), and the subsystem size increases in the horizontal direction ($l$).  The first four steps are shown.}
\label{fig:points}
\end{figure}

The largest discrepancies between the fit and the data appear at small $N_{s}$, with discrepancies remaining on the boundary (small $l$ and $l \approx N_{s}$) and lesser error near the center of the chain at larger values of $N_{s}$.  We then considered the removal of points from the boundary (points from small $l$ and $l \approx N_{s}$) in the pattern shown in Fig \ref{fig:points}.  When fitting to the entire data set we first remove a single point from the left and right (1 and $N_s - 1$) and perform a fit.  Starting over, we then remove a single point from $N_{s} = 4$ on both sides, but two points from all larger $N_s$ and perform a fit. Starting yet again we remove a single point from both sides for $N_{s} = 4$, two points from both sides for $N_{s} = 6$, and three points from either side for all larger $N_{s}$ and perform a fit.  We continue this until only the center points from $N_s/2$ remain.  This allows us to identify the contributions to the error from the points away from the center of the chain.

\begin{figure}
    \includegraphics[width=7.9cm]{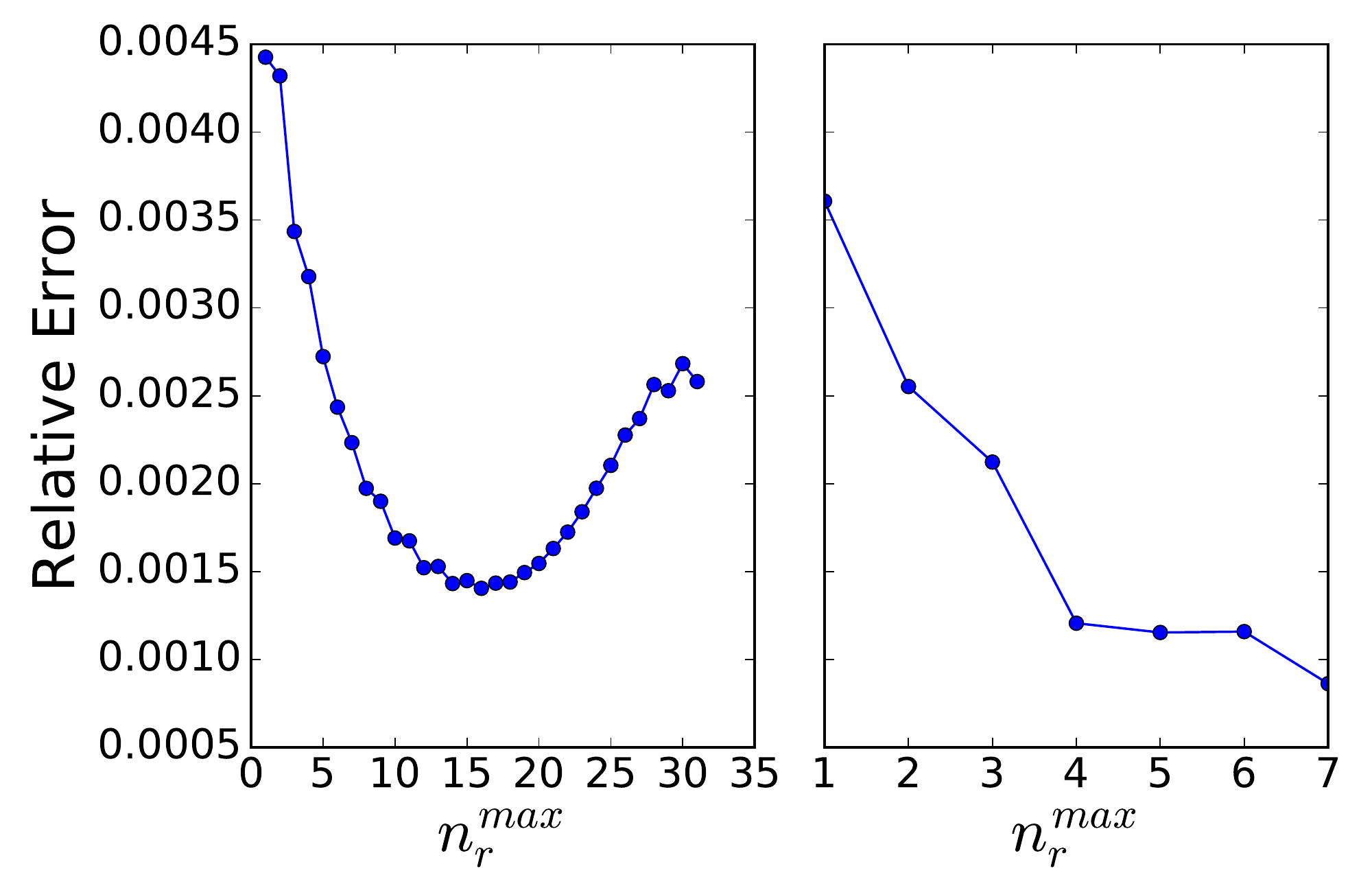}
    \caption{The average relative error versus the maximum number of points removed ($n_{r}^{max}$) from the boundary during the fit.  (left) Fitting to the entire data set, $4 \leq N_{s} \leq 64$, we see the optimum fit occurs when the majority of the smaller $N_s$ data has been removed except for the center points.  However a ``fan'' of data remains for the larger $N_s$ data.  (right) Fitting data such that $4 \leq N_{s} \leq 16$.  From this we see the optimum fit occurs particularly at $l = N_{s}/2$.}
    \label{fig:chi2-64-16}
\end{figure}

In order to measure the error on the fits we used the average relative error
\begin{equation}
	(\text{Relative Error})^{2} = \frac{1}{N} \sum_{i=0}^{N-1} \left( \frac{y_{i} - f(x_{i})}{y_{i}} \right)^{2}
\end{equation}
for $N$ data points, with $y_{i}$ the dependent data, $x_{i}$ the independent data, and $f$ the fit.  This measure has the advantage of being dimensionless.
The average relative error for two examples of $N_{s}=64$ and 16 is found in Fig. \ref{fig:chi2-64-16}.
In this Figure we see for small $N_{s}$ (right) that as the maximum number of points removed following the procedure above, $n_{r}^{max}$, approaches $N_{s}/2 - 1$ the relative error takes a minimum. We see the optimum fit for the entire data set (left) takes place when all points except the center ones have been removed for small $N_s$, and a ``fan'' of points remain for larger $N_{s}$.
% This is in line with Fig. \ref{fig:allpointsfit}.
When considering fits with $f$ and $f' \neq 0$ we again tried various, typical functional forms.  These included corrections $\propto 1/N_{s}$, $1/N_{s}^{2}$, and $1/\ln(N_{s})$ as well as corrections from \cite{PhysRevB.85.024418}.  All of these corrections obtained similar relative errors.

\begin{figure}
	\includegraphics[width=8.2cm]{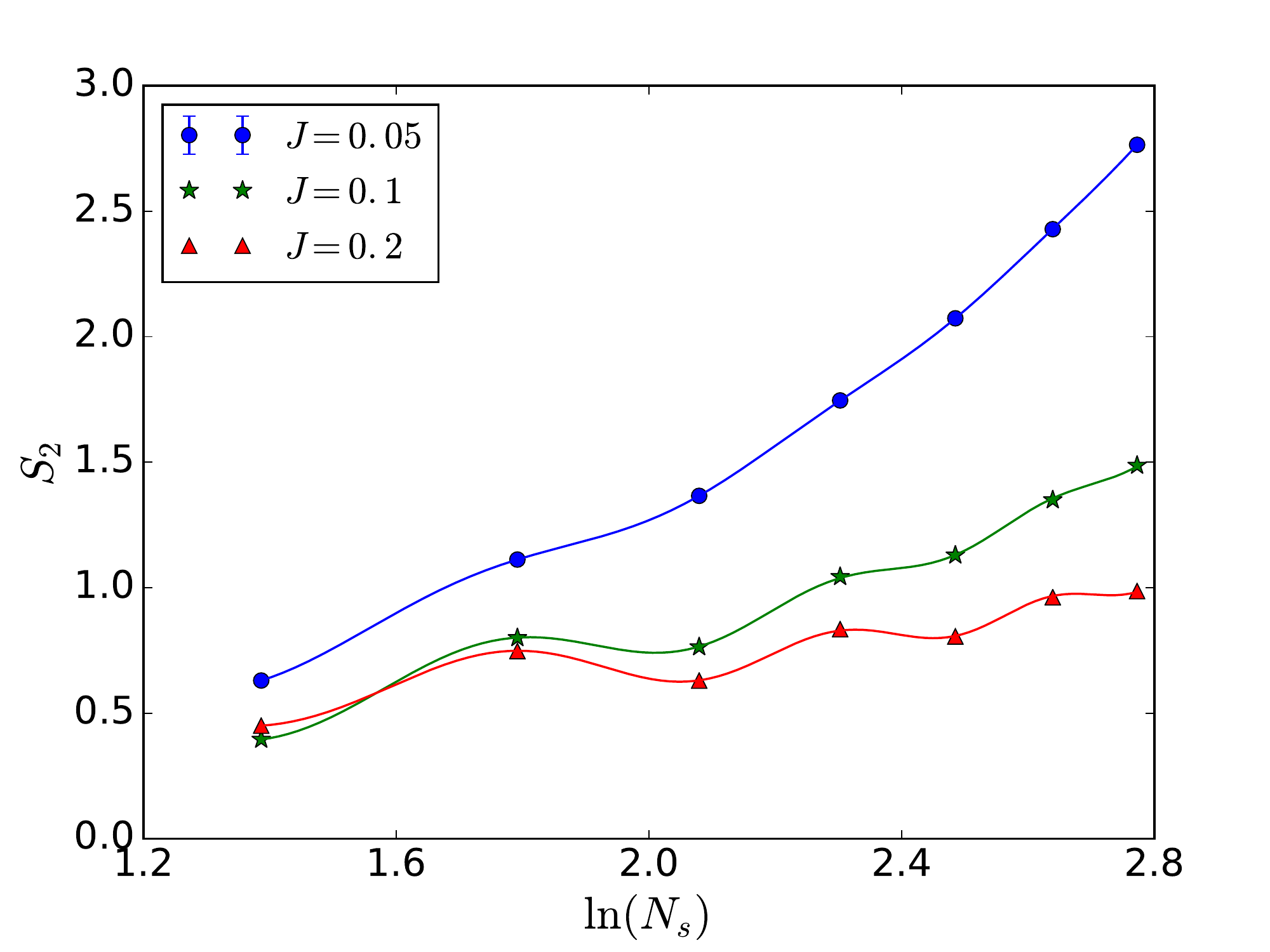}
    \caption{The second order R\'{e}nyi entropy, $S_{2}$, as a function of the logarithm of system size with a finite temperature and $l = N_{s}/2$.  Here $U = 1$ and $T = 0.04$ is shown.  The lines through the data points are fits using Eq. \ref{eq:cft-fit} with one addition term linear in the system size. The $J = 0.05$ data used sampling methods however the errors are smaller than the markers.}
    \label{fig:finiteTj0p2}
\end{figure}

We considered the effects of a finite temperature for a few cases on the R\'{e}nyi entropy.  An example of the finite temperature effects on the scaling with the system size is shown in Fig. \ref{fig:finiteTj0p2}.
As one can see the finite temperature effects are much more pronounced for smaller $J/U$.
To take the finite temperature effects into account it is enough to add a term linear in $l$ or $N_{s}$, for the subsystem or system scaling respectively.
The fits in Fig. \ref{fig:finiteTj0p2} and those used in the main text were done similarly, using Eq. \ref{eq:cft-fit} with an additional linear term.

It is possible to investigate the influence of statistical fluctuations of the pure $T=0$, $S_{2}$ data.  This can be accomplished by adding Gaussian fluctuations drawn from a distribution with a specified $\sigma_{S_{2}}$.  This synthetic data mimics the actual experimental results under the assumption that the error associated with each data point is the same regardless of system size or subsystem size.  By running many ``experiments'' one can see, approximately, how many measurements are needed for a given single experiment to obtain acceptable results.

\begin{figure}
    \includegraphics[width=7.9cm]{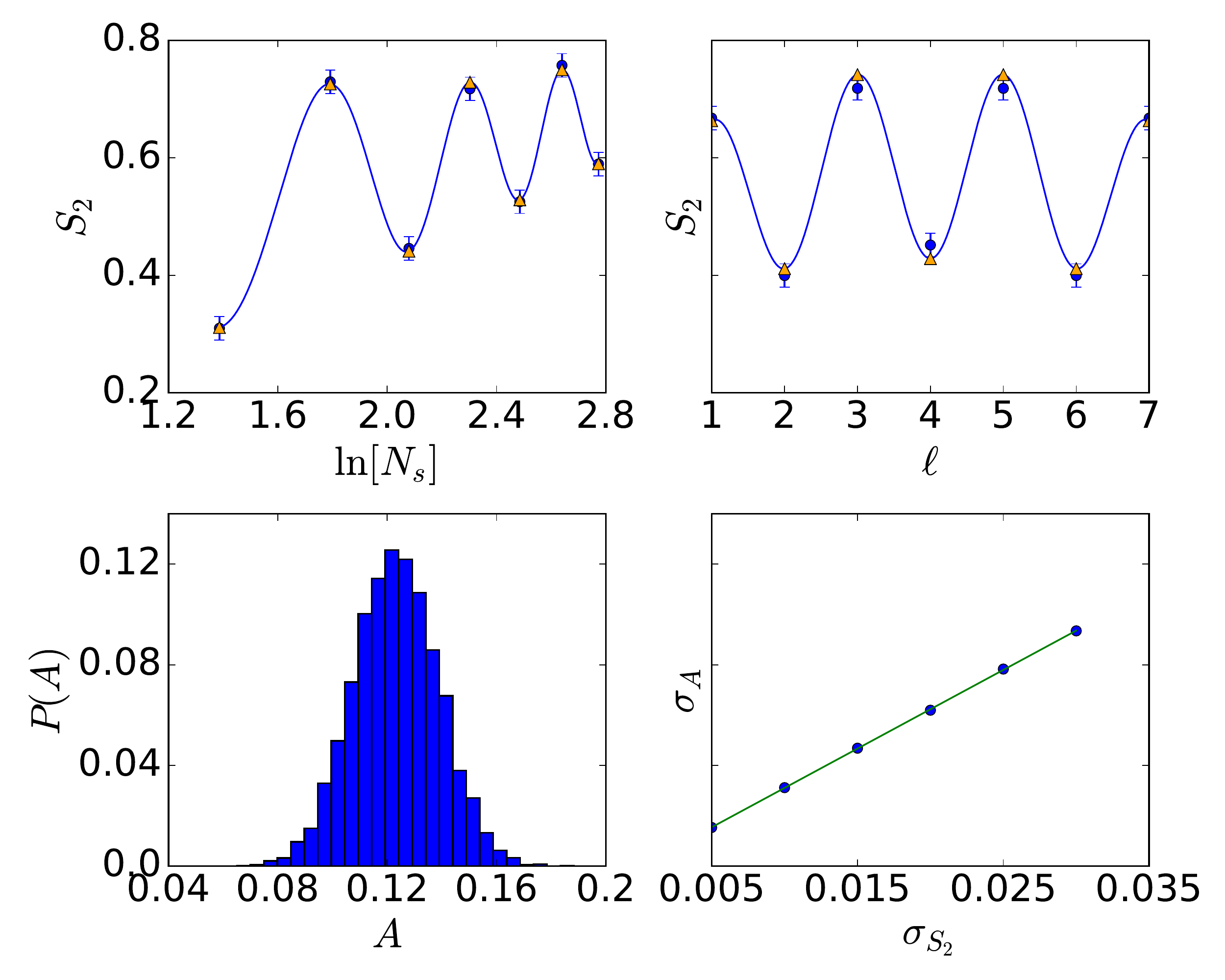}
    \caption{(top-left) $S_{2}$ with open boundary conditions at $T=0$ (orange triangles online) and the same data with added Gaussian fluctuations (blue circles online) as a function of $\ln[N_{s}]$ with a fit to the fluctuating data. (top-right) $S_{2}$ with open boundary conditions as a function of subsystem size for a fixed system size of $N_{s}=8$. (bottom-left) The probability distribution for $A$ values extracted from fits to fluctuating data. This distribution was built from 10,000 ``experiments''. (bottom-right) The error on the mean $A$ values extracted from fits to the fluctuating data versus the error on $S_{2}$.  The blue circles are the data, while the solid line is a linear fit with slope $\approx 3.1$.}
    \label{fig:gauss-flucs}
\end{figure}

To do this, one takes the pure, $T=0$ data and adds Gaussian noise with $\sigma_{S_{2}}$ to the data and then fits the data to the desired functional fit.  One does this many times and bins the fit parameters to obtain a histogram of the fit parameters for many experiments.  With the histogram one can extract $\sigma_{A}$, the error on the measurement of the central charge.  In addition one can establish the relationship between $\sigma_{S_{2}}$ and $\sigma_{A}$ to understand the size of the errors necessary on $S_{2}$ to obtain reliable estimates of $c$.  In Fig. \ref{fig:gauss-flucs} one can see how the addition of Gaussian fluctuations modifies the data and the estimates of $c$.
Using the distribution of $A$ values obtained from the fits we can extract $\sigma_{A}$.  In addition we can repeat this again for a different $\sigma_{S_{2}}$.  By doing this we find an approximately linear relationship between $\sigma_{A}$ and $\sigma_{S_{2}}$, such that $\sigma_{A} \approx 3.1 \sigma_{S_{2}}$.
For a systematic error of 0.02 on the measurement of $A$, the uncertainty in $S_{2}$ must be approximately 0.006.  This gives an estimate on the number of measurements necessary, assuming a maximum entropy of 1.1.  Using Eq. (6) in the main text we find on the order of 193,000 measurements.

\end{document}